\documentclass[letterpaper,twocolumn,10pt]{article}

\usepackage{mathptmx}
\usepackage{usenix,endnotes}
\usepackage{amsmath}
\usepackage{amssymb}
\usepackage{amsfonts}
\usepackage{epsfig}
\usepackage{subfigure}
\usepackage{calc}
\usepackage{color}
\usepackage[all]{xy}
\usepackage{bm}
\usepackage{algorithmicx}
 \usepackage{algpseudocode}
\usepackage{algorithm}
\usepackage{enumerate}
\usepackage{hyperref}
\usepackage{fancybox}
\usepackage{titling}

\newtheorem{defn}{Definition}
\newtheorem{thm}{Theorem}[section]
\newtheorem{cor}[thm]{Corollary}
\newtheorem{prop}{Proposition}

\newtheorem{lem}[thm]{Lemma}
\newtheorem{conj}[thm]{Conjecture}
\newtheorem{constr}[thm]{Construction}
\newtheorem{note}{Remark}

\newcommand{\bit}{\begin{itemize}}
\newcommand{\eit}{\end{itemize}}
\newcommand{\bcor}{\begin{cor}}
\newcommand{\ecor}{\end{cor}}
\newcommand{\beq}{\begin{equation}}
\newcommand{\eeq}{\end{equation}}
\newcommand{\beqn}{\begin{equation*}}
\newcommand{\eeqn}{\end{equation*}}
\newcommand{\bea}{\begin{eqnarray}}
\newcommand{\eea}{\end{eqnarray}}
\newcommand{\bean}{\begin{eqnarray*}}
\newcommand{\eean}{\end{eqnarray*}}
\newcommand{\ben}{\begin{enumerate}}
\newcommand{\een}{\end{enumerate}}
\newcommand{\bdefn}{\begin{defn}}
\newcommand{\edefn}{\end{defn}}
\newcommand{\bnote}{\begin{note}}
\newcommand{\enote}{\end{note}}
\newcommand{\bprop}{\begin{prop}}
\newcommand{\eprop}{\end{prop}}
\newcommand{\blem}{\begin{lem}}
\newcommand{\elem}{\end{lem}}
\newcommand{\bthm}{\begin{thm}}
\newcommand{\ethm}{\end{thm}}
\newcommand{\bconj}{\begin{conj}}
\newcommand{\econj}{\end{conj}}
\newcommand{\bconstr}{\begin{constr}}
\newcommand{\econstr}{\end{constr}}
\newcommand{\bpf}{\begin{proof}}
\newcommand{\epf}{\end{proof}}

\let\oldbibliography\thebibliography
\renewcommand{\thebibliography}[1]{%
 \oldbibliography{#1}%
 \setlength{\parskip}{0pt}%
 \setlength{\itemsep}{0pt plus 0.3ex}
}

\title{Evaluation of Codes with Inherent Double Replication for Hadoop}
\date{}
\author{\textnormal{M. Nikhil Krishnan, N. Prakash, V. Lalitha,} \\ \textnormal{Birenjith Sasidharan, P. Vijay Kumar} \\ \textnormal{Indian Institute of Science, Bangalore}  \and  \textnormal{Srinivasan Narayanamurthy,} \\ \textnormal{Ranjit Kumar,  Siddhartha Nandi}  \\ \textnormal{NetApp Inc.}}
        
\begin{document}

\maketitle

\begin{abstract}

In this paper, we evaluate the efficacy, in a Hadoop setting,
of two coding schemes, both possessing an inherent
double replication of data. The two coding schemes 
belong to the class of regenerating and
locally regenerating codes respectively, and these two classes are representative of recent
advances made in designing codes for the efficient storage of data in a distributed setting.  
In comparison with triple replication, double replication permits a significant reduction in storage overhead,
while delivering good MapReduce performance under moderate work loads. The 
two coding solutions under evaluation here, add only moderately to the storage
overhead of double replication, while simultaneously offering 
reliability levels similar to that of triple replication.   

One might expect from the property of inherent data duplication that the performance of these codes in executing a MapReduce job would be comparable to that of double replication.  However, a second feature of this class of code comes into play here, namely that under both coding schemes analyzed here, multiple blocks from the same coded stripe are required to be stored on the same node.   This concentration of data belonging to a single stripe negatively impacts MapReduce execution times.  However, much of this effect can be undone by simply adding a larger number of processors per node.  Further improvements are possible if one tailors the Map task scheduler to the codes under consideration.   We present both experimental and simulation results that validate these observations.  

\end{abstract}

\vspace{-0.1in}

\section{Introduction}
\vspace{-0.1in}

Hadoop~\cite{hadoop} is an open-source platform dealing with distributed storage whose file system is known as the Hadoop distributed file system (HDFS).  The primary objective is to store a
collection of files in such a way that distributed computation on the stored data can be carried out efficiently under the MapReduce (MR)~\cite{mapreduce} paradigm. Any file that needs to be stored is divided into blocks, typically of size
$64-256$ MB and these blocks are replicated and are stored across the distributed storage network, such
that no two replicas of the same block are stored in the same storage node. 

Replication of data, in addition to providing resiliency against irrecoverable data loss, supports efficient MR computation in two ways. Firstly, it ensures {\it availability} of data during transient node failures. Such failures, which do not cause data loss, are the norm~\cite{ford2010availability} in large-scale storage systems, and hence minimizing the number of repairs carried out to handle transient failures, can result in significant savings in network bandwidth~\cite{sathiamoorthy}. Secondly, replication helps to increase {\it data locality} for MR computation. A Hadoop node can typically perform $2$ to $8$ map tasks in parallel, depending on the number of processor cores available to it. As a result, in a system which is expected to handle multiple compute jobs simultaneously, the presence of replicas will increase the chance that any given map task can be assigned to  a node which contains the data block required by the task.  Such a task is called a {\it local} task. A non-local or {\it remote} task needs data to be fetched across the network to the node where the task is to be executed, leading to increased delay in job execution as well as increased network bandwidth usage~\cite{mapreduce}.

Triple replication is often used in a Hadoop system as it provides the desired level of resiliency as well as availability of data.  In comparison with triple replication, while double replication permits a significant reduction in storage overhead, and can potentially deliver good MapReduce performance under moderate work loads, it leaves the system vulnerable to irrecoverable data loss in the event of some patterns of two-node failure.  In an effort to avoid the large overheads associated with triple replication, storage-efficient erasure codes such as Reed-Solomon (RS) codes and some variants have recently been employed in Facebook's Hadoop clusters~\cite{hdfs_raid}, \cite{sathiamoorthy}. These codes store only a single copy of the data file and use parity blocks, as in RAID, to offer protection against failures. Application of these codes is thus limited to the storage of cold data, i.e., data on which MR jobs are rarely carried out.

In this paper, we evaluate the efficacy, in a Hadoop setting, of two coding schemes, both possessing an inherent
double replication of data. The two coding schemes are described in Section \ref{sec:pentagon}. Performance analysis of metrics like storage overhead and mean time to data loss (MTTDL), and also simulation results showing data locality for map tasks are presented in Section \ref{sec:analysis}. Details of the experimental set-up used for MR performance evaluation, and the associated findings can be found in Section \ref{sec:experiments}. Conclusions and future work are presented in Section \ref{sec:future} and an overview of the related literature is presented in  Section \ref{sec:related}.

\vspace{-0.1in}
\section{The Two Coding Schemes} \label{sec:pentagon}

\vspace{-0.1in}
\subsection{Pentagon Code}

\begin{figure}[h]
  \centering
  \subfigure[Pentagon Code]{\label{fig:pentagon}\includegraphics[height=1.6in]{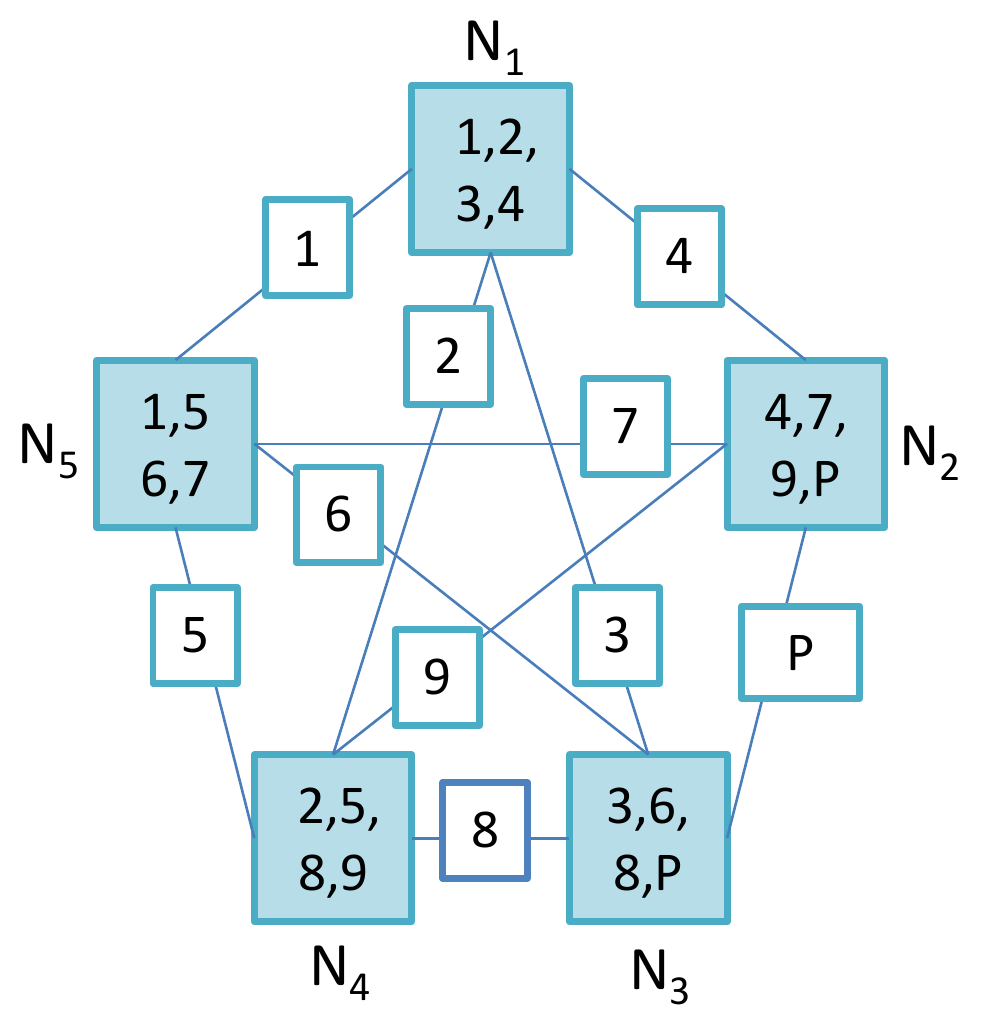}}
  \hspace{0.05in}
  \subfigure[Heptagon-Local Code]{\label{fig:heptagon_local}\includegraphics[height=0.9in]{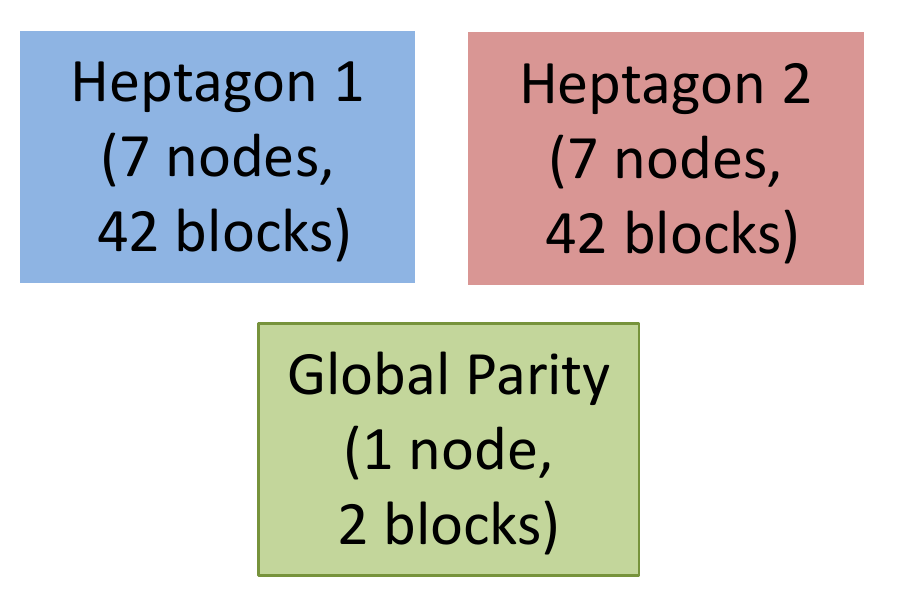}}
  \vspace{-0.1in}
  \caption{Codes with inherent double replication of data.}
  \label{fig:regen_framework}
\end{figure}

The first code which we refer to here as the {\it pentagon code}, is a simple instance of a family of codes known as repair-by-transfer minimum-bandwidth regenerating codes~\cite{ShaRasKumRam_rbt}.   In the pentagon code, $9$ data blocks are encoded\footnote{In case, the file to be encoded is larger than $9$ blocks, the file is striped into sets of $9$ blocks each and each stripe is separately encoded.} into $20$ coded blocks and stored in $5$ nodes with $4$ blocks assigned to each node. Given the $9$ data blocks, a $10^{\text{th}}$ block which is an XOR parity of these $9$ blocks is first computed; the resulting $10$ blocks are replicated once more and stored in $5$ nodes as shown in Fig. \ref{fig:pentagon}. To explain the distribution of the blocks among the $5$ nodes, consider a fully connected graph with the $5$ target nodes as its vertices. Note that there are ${5 \choose 2} = 10$ edges in this graph.  Place each of the $10$ distinct blocks on the $10$ edges of this graph. The blocks stored in a node are then simply those associated to the edges that are incident on the particular node.  An important point to note here is that the pentagon code stores multiple blocks of a coded stripe in the same node. Such codes, in general, are known as array codes.

It can be readily verified that the contents of any $3$ nodes suffice to recover all $9$ data blocks and
thus the code is resilient to $2$-node failure. Single-node repair is accomplished, simply by transferring to the replacement of the failed node, the blocks shared in common by the failed node with each of the remaining $4$ nodes. In the case of two node failures, nodes $N_1$ and $N_2$ say, $8$ blocks have to be recovered in all, out of which $6$ can be recovered just by copying the replicas of these blocks present in the remaining $3$ nodes. For the recovery of block $4$, nodes $N_3$, $N_4$ and $N_5$ internally compute the {\it partial parities} $P_3 = 3+6+P$, $P_4 = 2+8+9$ and $P_5 = 1+5+7$, respectively and transfer these partial parities to the first replacement node.  The replacement node computes $P_3 + P_4 + P_5$ and recovers block $4$, which is then copied to the second replacement node as well. Thus, the overall network data transfer incurred in repairing the two nodes (also known as repair bandwidth) is $10$ blocks. The repair bandwidth savings arising from the usage of partial parties is an intrinsic advantage possessed by array codes.

The construction of the pentagon code, suggests that a third coding scheme be included for comparison. Under this scheme, given $9$ data blocks, one computes a $10^{\text{th}}$ parity block as above and then duplicates each coded block to obtain a total of $20$ blocks.  These $20$ blocks are then stored across $20$ different nodes of the network.  This scheme is termed in the literature~\cite{xin2003reliability} as a RAID+mirroring scheme.  The code itself is termed as a $(10,9)$ RAID+m code.  The performance comparison we present in Section \ref{sec:analysis}, includes comparison with the RAID+m scheme.


\vspace{-0.1in}
\subsection{Heptagon-Local Code}

The heptagon-local code is built on top of a heptagon-code, which is the analogue of the pentagon code for $7$ nodes.  The heptagon code encodes $20$ data blocks into $42$ blocks and stores them in $7$ nodes, with each node hosting $6$ blocks following a placement rule similar to that applied in the case of the pentagon code. The storage overhead of the heptagon code is less than that of the pentagon code;  however it has a lower level of resiliency (see Table \ref{tab:bw_compare}). 

The heptagon-local code, is an instance of a family of codes known as locally regenerating codes~\cite{KamPraLalKum}.  This family presents a simple way to increase the resiliency of the heptagon code, at the cost of a moderate increase in storage overhead. In this code, $40$ data blocks are encoded into $86$ blocks and stored in $15$ nodes, as follows: (1) The $40$ data blocks are first split into two sets of $20$ each, which are then individually encoded by two instances of the heptagon code taking care to ensure that the $7$ nodes chosen for the first heptagon are distinct from those chosen for the second heptagon. (2) Two \textit{global parity blocks} are computed as functions of all $40$ data blocks, and are placed in an additional {\em global parity node}. This computation involves Galois field arithmetic as in the case of RAID-6. In this code, the individual heptagon codes themselves are referred to as \textit{local codes}. In a rack-aware HDFS implementation, the two heptagons and the global parity node would be placed in three different racks. 

The heptagon-local code can recover from any pattern of $3$ node erasures.  The failure of $1$ or $2$ nodes lying within a heptagon, can be handled \textit{locally}, i.e., by accessing other nodes within the same heptagon. In the case where $3$ nodes belonging to a single heptagon fail, the recovery will involve accessing the contents of the second heptagon as well as the contents of the global parity node. Here again, the repair bandwidth can be reduced by making use of partial parities.

\vspace{-0.1in}
\section{Performance Analysis for Hadoop} \label{sec:analysis}

\vspace{-0.1in}
\subsection{File System Performance}

A comparison of performance metrics is presented in Table \ref{tab:bw_compare}. The MTTDL values shown are computed assuming a $25$ node system, using standard node failure and repair models available in the literature~\cite{xin2003reliability}. 

\begin{table}[h!]
\centering
\begin{tabular}{||c|c|c|c|c||}
\hline
Code & Storage & Code  &  MTTDL\\
&   Overhead & Length &  (in years)    \\
\hline
\hline
3-rep & 3x & 3&    1.20e+09 \\
\hline
pentagon & 2.22x &  5 &  1.05e+08 \\
\hline
heptagon & 2.1x & 7 &  2.68e+07 \\
\hline
heptagon-local & 2.15x & 15 &  8.34e+09 \\
\hline
$(10,9)$ RAID$+$m & 2.22x  &  20 & 2.03e+09 \\
\hline
$(12,11)$ RAID$+$m & 2.18x  &  24 & 6.50e+08 \\
\hline
\end{tabular}
\caption{Comparing storage overhead, code length and MTTDL of various coding schemes.}
\label{tab:bw_compare}
\end{table}

An important advantage of the proposed codes is the smaller value of their code length for a given value of storage overhead, in comparison with the corresponding RAID+m scheme. The code length specifies the number of data nodes over which a coded stripe is distributed.  For instance, both the pentagon and the $(10, 9)$ RAID+m code have a storage overhead of $2.22$; clearly between the two codes, only the pentagon code is feasible in a Hadoop system possessing just $20$ nodes. The length of the RAID+m solution can be decreased below $20$ only at the expense of increased storage overhead. For larger systems, the heptagon-local code turns out to be an attractive choice in terms of all three performance metrics. 


The pentagon and heptagon-local codes also are efficient in terms of the repair-bandwidth needed during \text{on-the-fly} repair of a lost data block, during an MR computation. Imagine a situation where the two nodes which store the two replicas of a particular block are temporarily down, and a map task is initiated on this particular block. While the $(10, 9)$ RAID+m solution needs a repair bandwidth of $9$ blocks, a repair bandwidth of $3$ blocks suffices in the case of the pentagon code, arising from the pentagon code's ability to compute and make use of partial parities. Typically, in Hadoop, repair jobs themselves are performed as MR jobs and in such a case, computation of the partial parities can be incorporated into the MR-job through the use of ``combine" functions\footnote{Combine function is a mechanism, while performing MR jobs, to consolidate the outputs of the various map tasks which are performed on the same Hadoop node.}.

\vspace{-0.1in}
\subsection{Data Locality of Map Tasks}

We evaluate here the data locality under Map tasks of the pentagon and heptagon local codes in a moderately-loaded Hadoop system.  We quantify the notion of load as follows (with an example):  A $100$-node system that handles $250$ map tasks, with $4$ map slots per node, is said to be operating under a load of $\frac{250}{(4 \times 100)} \times 100 \ = \ 62.5\%$. By moderately loaded, we mean that the system is most of the time, operating on a $\leq100\%$ load. 

\begin{figure}[h!]
\begin{center}
\includegraphics[width=3in]{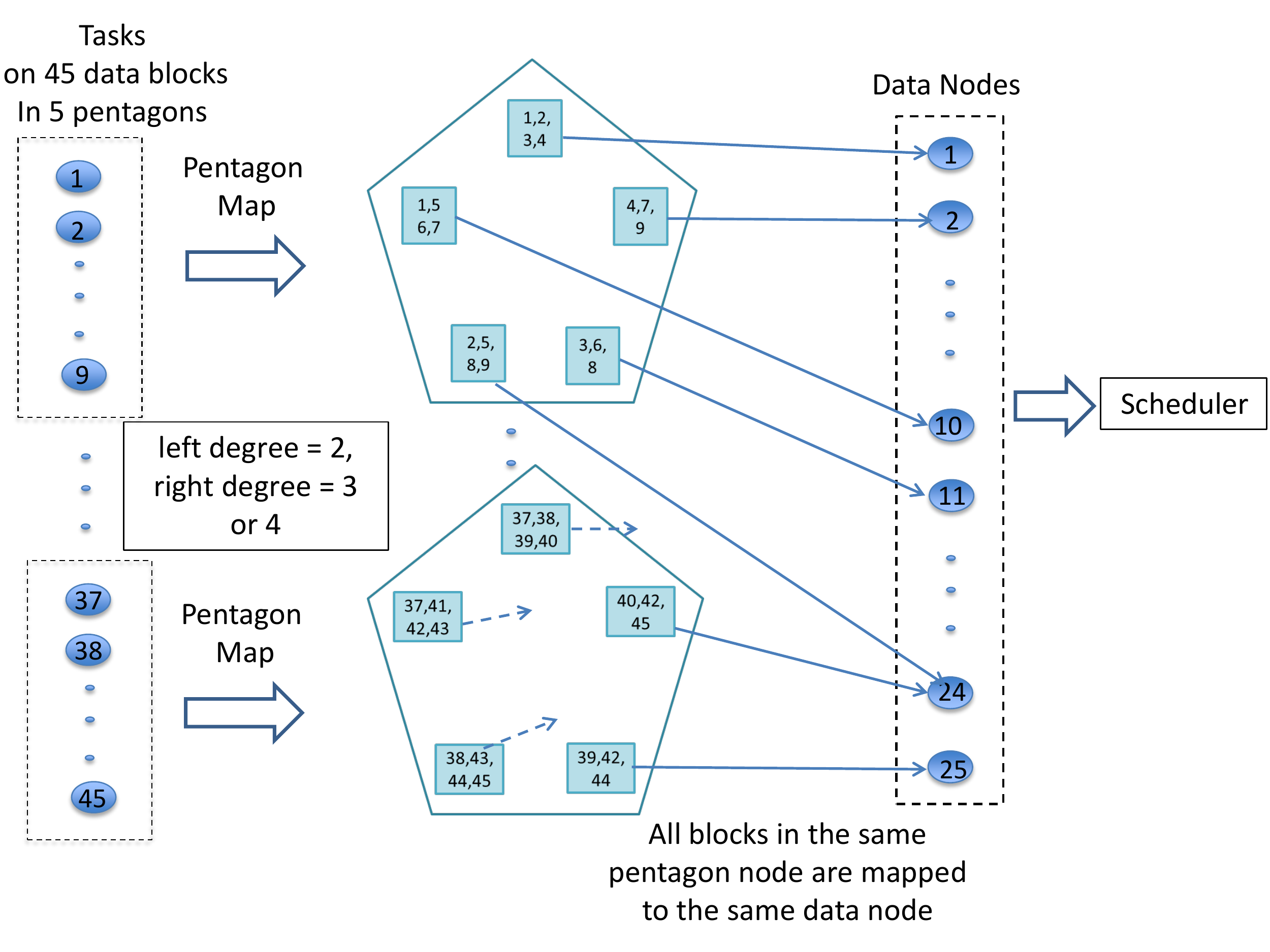}
\end{center}
\vspace{-0.25in}
\caption{The manner in which tasks are mapped onto nodes in the case of the pentagon code is shown here.}
\label{fig:pta}
\end{figure}

The map-task-assignment problem can be modeled as a maximum-matching problem on a bipartite graph, with the tasks on one side and the nodes on the other. The edges on this graph indicate the nodes where the replicas of the blocks reside. We note that the choice of a particular coding solution essentially determines the manner in which the edges are incident on the vertices lying to the right, i.e., the nodes (see Fig.~\ref{fig:pta}).  From a practical point of view, maximum-matching algorithms are computationally intensive. Hadoop uses instead, a simple algorithm called delay scheduling for task assignment~\cite{zaharia2010delay}.

\begin{figure}[h]
\begin{center}
\includegraphics[width=3.0in]{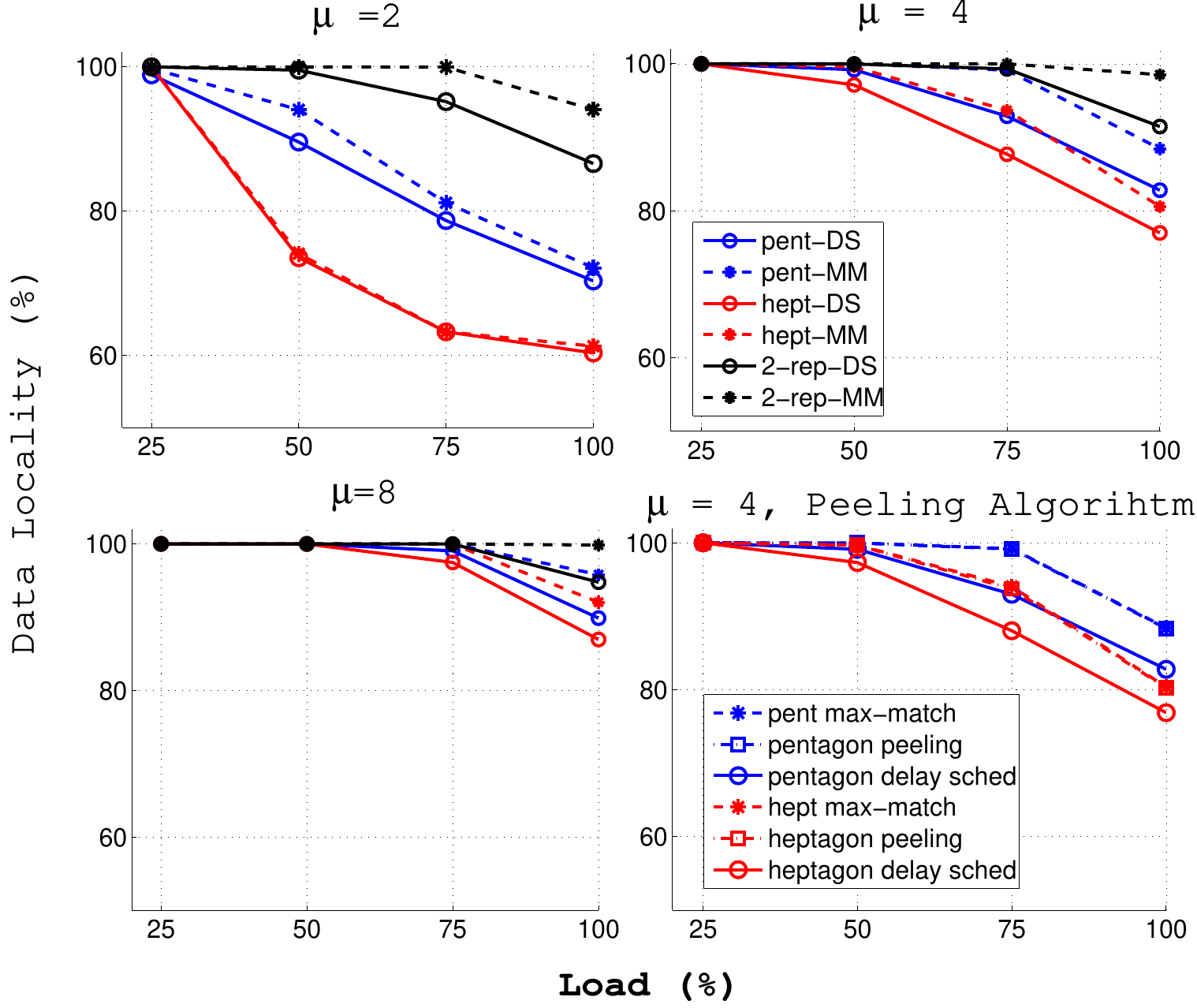}
\end{center}
\vspace{-0.25in}
\caption{Percentage data locality with $2$-rep, pentagon and heptagon-local codes, as a function of job load on a system, for the cases of $\mu = 2,4,8$ map slots per node. Also shown, is the improvement in locality obtained over the delay scheduler via the modified peeling algorithm, for the case of $\mu =4$.}
\label{fig:locality}
\end{figure}

To compare the data locality of the various codes, we simulated the delay-scheduling algorithm in a $25$-node system, under various load conditions. The maximum-matching algorithm was also simulated as a benchmark. Simulation results are presented for the $2$-rep, pentagon and the heptagon codes, for the cases of $2$, $4$ and $8$ map slots per node (see Fig. \ref{fig:locality}). We note that the data locality of the heptagon-local code will be similar to that of the heptagon code, since the global parity node does not play a role in task assignment. Also, the locality of the $2$-rep systems is indicative of the locality of any of the RAID+m solutions. 

\vspace{-0.15in}
We see from the plots that there is a significant loss in data locality with $2$ map slots per node, for the proposed coding schemes, with respect to double replication. This is an artifact of the fact that under both coding schemes analyzed here, multiple blocks from the same coded stripe are required to be stored on the same node (see Fig.~\ref{fig:pta}).   This concentration of data belonging to a single stripe negatively impacts MapReduce execution times.    Also, since the heptagon code has a greater concentration of data blocks in comparison with the pentagon code, it suffers more in this respect. However, as seen from the simulations, the loss in locality decreases with increasing number of map slots per node. For instance, both the pentagon and heptagon-local codes have locality greater than $90 \%$ at $100 \%$ load, with $8$ map slots. The plots also suggest that there is scope for improvement in locality by using task assignment algorithms, other than the delay scheduler. 

A simple algorithm known as the {\em peeling algorithm} was proposed in \cite{xie2012degree}, for task-assignment problems. Appropriate modifications to this algorithm allows it to be used in pentagon or heptagon-coded Hadoop systems. A simulation using the modified peeling algorithm, for the case of $4$ map slots per node is also shown in Fig.~\ref{fig:locality}; the performance gains in locality are clearly evident.

\vspace{-0.1in}
\section{Experimental Set-up and Evaluation} \label{sec:experiments}
\vspace{-0.1in}
An implementation of the pentagon and the heptagon codes was carried out in HDFS, taking Facebook's open-source HDFS-RAID~\cite{hdfs_raid} module (hadoop-0.20) as the baseline software. The main challenge in the implementation was in handling the array-nature of these codes, as they necessitate the logical grouping of blocks within a node.  Preliminary experiments, to ascertain MR performance, were conducted on two different Hadoop systems, which were chosen to have different numbers of processor cores per Hadoop node.

\textit{Set-up $1$ (2 map slots)}: This set-up had $25$ data nodes and the hardware used for each of these nodes was a dual-core IBM laptop, having $3$ GB of RAM and $150$ GB of hard disk space. The Hadoop data block size was set to $128$ MB. Also, each node was configured with $2$ map and $1$ reduce slots. We tested the pentagon and the heptagon codes in this set-up.

\textit{Set-up $2$ (4 map slots)}: This had $9$ data nodes and each of these nodes was a server class machine having $4$ processor cores per node, $24$ GB of RAM and $2$ TB of hard disk space. The Hadoop data block size was set to $512$ MB. Also, each node was configured with $4$ map and $2$ reduce slots. We tested the pentagon code in this set-up.

In both the set-ups, an additional master node was used to host all the controllers namely NameNode, JobTracker and RaidNode. All machines ran Ubuntu $12.04$ for their operating system. Also, all nodes were configured to be part of a single rack and shared a private $10$ Gbps Ethernet LAN.

\vspace{-0.1in}
\subsection{MapReduce Performance}
\vspace{-0.1in}

The Terasort job was executed at various load points (from $25 \%$ to $100 \%$) and under various coding schemes. We calculate values of data locality, job execution time and network traffic during job execution, averaged over multiple runs. We use Hadoop's inbuilt delay-scheduling algorithm for map-task assignment, with the delay set such that every node has a chance to assign two (four) local map tasks in the first (second) set-up. Features such as cap-based load management and speculative execution were turned off.

\begin{figure}[h!]
\begin{center}
\includegraphics[width=3in]{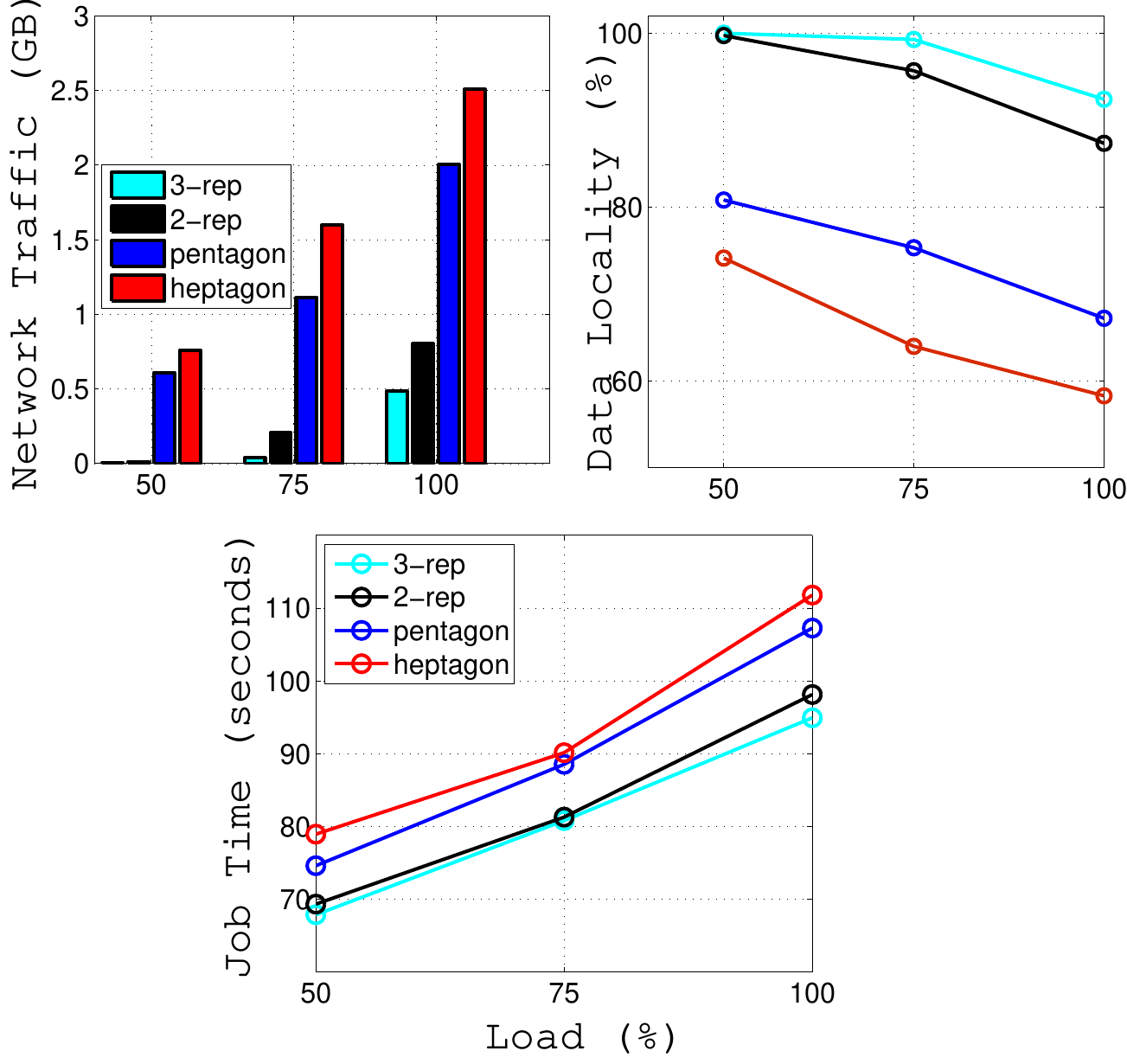}
\end{center}
\vspace{-0.25in}
\caption{Network traffic, data locality and job time for the Terasort job in set-up $1$, having $2$ map-slots per node.}
\label{fig:locality_setup1}
\end{figure}

\begin{figure}[h!]
\begin{center}
\includegraphics[width=3in]{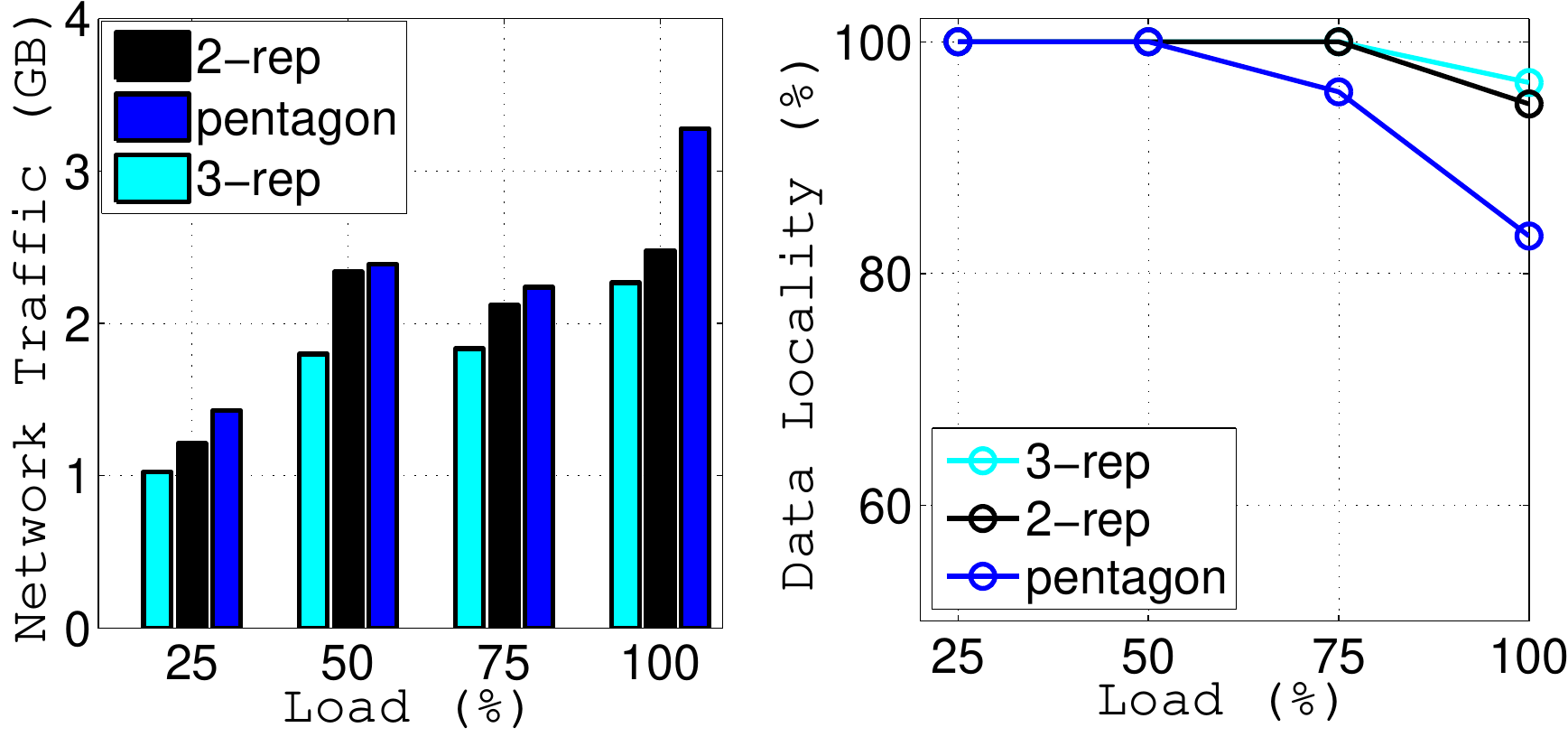}
\end{center}
\vspace{-0.25in}
\caption{Network traffic and data locality for the Terasort job in set-up $2$, having $4$ map-slots per node.}
\label{fig:locality_setup2}
\end{figure}

The measurements are shown in Fig. \ref{fig:locality_setup1} and \ref{fig:locality_setup2}, respectively for the cases of the first and second set-up.  The following conclusions can be drawn from the plots: $(i)$ At moderate loads, the performance of $2$-rep is very close to that of $3$-rep. $(ii)$ The data locality curves, in both the set-ups, exhibit the same trend as seen in the simulations curves in Fig. \ref{fig:locality}. $(iii)$ It can be verified that the excess network traffic for either the pentagon or the heptagon code with respect to $2$-rep is almost entirely due to the corresponding loss in data locality. $(iv)$ Also, as expected, there is a substantial loss in performance in the case of $2$ processor cores; however with $4$ cores, we see that the pentagon code has performance very close to that of the $2$-rep code even at a load of $75 \%$.

\section{Conclusions and Future Work} \label{sec:future}
\vspace{-0.1in}

In this work, we have implemented in Hadoop two coding schemes having inherent double replication of data, and have carried out a preliminary MR performance-evaluation under this coded system. Our measurements suggest that under moderate work loads, with an increased number of processors per node, the MR performance is comparable to that of double replication. In the next phase of our work, we plan to enhance the current implementation in two ways: $(i)$ Implement the heptagon-local code, and $(ii)$ implement the modified peeling algorithm, which was simulated (see Fig. \ref{fig:locality}), as an alternative to delay scheduler for map task assignment. We also plan to measure MR performance on a variety of work loads, for the enhanced system. Other important metrics, like encoding duration and MR performance in the presence of node failures (with the usage of partial parities) also need to be ascertained.

\vspace{-0.2in}
\section{Related Work}\label{sec:related}
\vspace{-0.15in}
Regenerating codes and locally repairable codes are respectively introduced in \cite{DimGodWuWaiRam} and \cite{GopHuaSimYek}. The authors of \cite{sathiamoorthy} describe an implementation of locally repairable codes in HDFS and evaluate the  savings in network traffic during node repairs, with these codes. A study of locally repairable codes in the context of Windows Azure storage can be found in \cite{HuaSimXu_etal_azure}, where it is shown that these codes are better choices than Reed-Solomon codes, in terms of their reliability vs storage overhead performance. Regenerating codes have been implemented in a multiple-cloud system in \cite{HuCheLeeTan}, and for HDFS in \cite{LiLinLee}. Both works focus on the problem of decreasing repair bandwidth. Statistics regarding node failures in Hadoop clusters with thousands of nodes have been reported in \cite{sathiamoorthy}, \cite{RasShaHotStorage}. A class of erasure codes for minimizing I/O during recovery and degraded reads with application to cloud-file systems has been proposed in \cite{khan2012rethinking}.
\vspace{-0.2in}

\section*{Acknowledgements} 
\vspace{-0.1in}

The research of P. Vijay Kumar is supported in part by the National Science Foundation under Grant 0964507 and in part by the NetApp Faculty Fellowship. The works of V. Lalitha and Birenjith Sasidharan are supported by a TCS Research Scholarship.

\vspace{-0.1in}

{\scriptsize
\bibliographystyle{IEEEtran}
\bibliography{FAST_bib}
}

\end{document}